\documentclass{PoS}
\pdfoutput=1
\usepackage{graphicx}
\usepackage{amsmath} 
\usepackage{amsfonts}% f^?ur Mengen-Symbole: \mathbb{ABC}
\usepackage{amssymb}
\usepackage{subfigure}
\newcommand{\Trace}{\operatorname{Tr}}

\title{Comparing the vacuum structure of quenched and dynamical configurations}

\ShortTitle{Comparing the vacuum structure of quenched and dynamical configurations}

\author{ Falk Bruckmann, \speaker{Florian Gruber} and  Andreas Sch\"afer\\
        University of Regensburg\\
        E-mail: \email{falk.bruckmann@physik.uni-regensburg.de},\\
                \email{florian.gruber@physik.uni-regensburg.de},\\
		\email{andreas.schaefer@physik.uni-regensburg.de}}

\abstract{
We systematically compare filtering methods used to extract topological structures
on $SU(3)$ lattice configurations. We show that there is a strong correlation of the topological charge densities obtained by APE and Stout smearing. To get rid of artifacts of these methods, we analyze structures that are also seen by Laplace filtering.
This combined analysis shows that the topological charge density is more fragmented in the presence of dynamical quarks.
}

\FullConference{The XXVII International Symposium on Lattice Field Theory\\
		 July 26-31, 2009\\
		 Peking University, Beijing, China}

\begin{document}

\section{Filtering Methods and Topological Charge Density}
Many methods have been developed to extract the IR content from lattice data. Unfortunately, all these methods introduce ambiguities and parameters. Thus, to get a coherent picture of the topological structure of the QCD vacuum, it is necessary to find ways of controlling or even removing these ambiguities.

One of the first attempts to filter out the UV `noise' has been \emph{APE smearing} \cite{APE}, defined as:
\begin{equation}
\text{\includegraphics[width=0.48\textwidth]{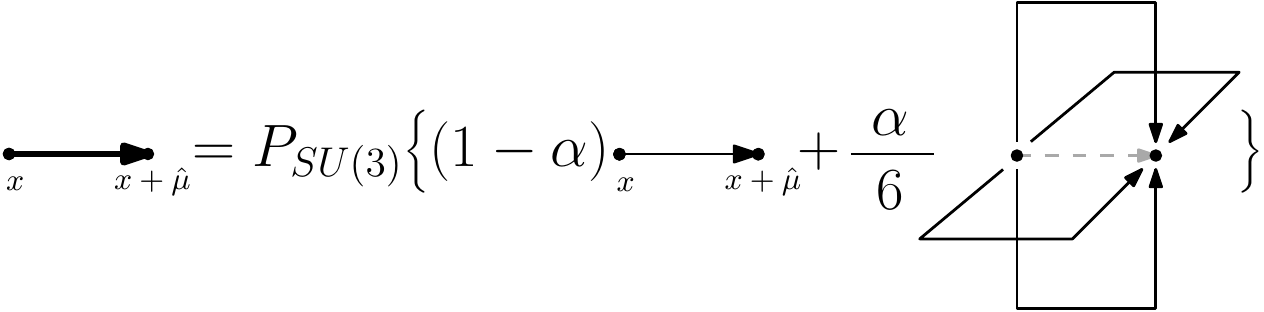}}
\end{equation}
where $\alpha$ determines the weight of the old link and the sum of the attached staples. Throughout this work we have used the standard value $\alpha=0.45$. The right hand side has to be projected back to the gauge group. The projection onto $SU(3)$ is not unique, but a common choice is to define $P_{SU(3)}(W)$ as the that element of the group $V \in SU(3)$ that maximizes $\operatorname{Re}\Trace\{VW^{\dagger}\}$
\footnote{The maximum is found iteratively using code included in the CHROMA software package for lattice QCD \cite{Edwards2005}.}.
\emph{Stout smearing} \cite{Morningstar2004a} avoids this projection by using the exponential map
 $U_\mu^{ {\rm Stout}}\equiv \exp\{i\; Q_{\mu}(U ,\rho)\}\cdot U_\mu^{{\rm old}}$ ,
where
\begin{equation}
\text{\includegraphics[width=0.43\textwidth]{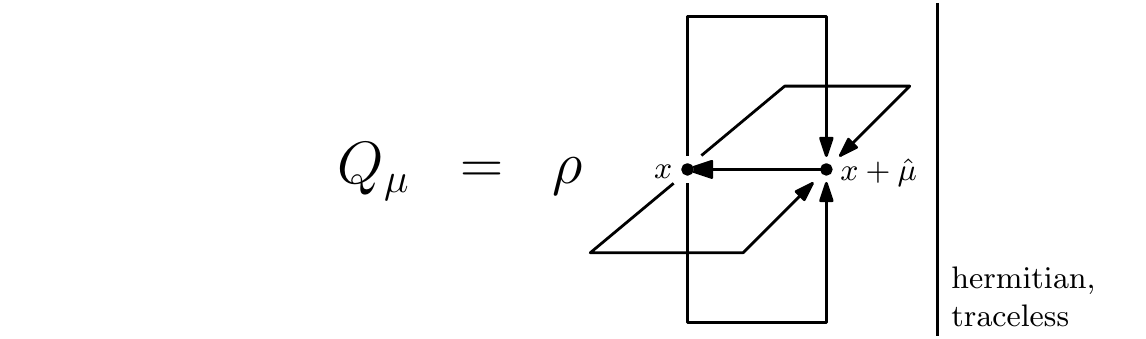}}
\end{equation}
is a hermitian traceless matrix constructed from all plaquettes containing the old link $U_\mu^{\rm old}$ and a smearing parameter $\rho$ and hence $e^{iQ} \in SU(3)$.

Both smearing methods are iterative procedures, which lead to smoother configurations in each step. But it is hard to decide how many smearing steps one can apply without loosing information about the underlying structure. 

A relatively new method is \emph{Laplace filtering} \cite{Bruckmann2005a}. This method is based on a truncated spectral decomposition of the links in terms of eigenmodes $\Phi_n(x)$ of the covariant lattice Laplacian:

\begin{equation}
 U_\mu^{\text{Laplace}}(x) = P_{SU(N_c)}\left\{-\sum_{n=1}^{N}\lambda_{n}\Phi_n(x)\otimes\Phi_n^{\dagger}(x+\hat\mu)\right\}.
\end{equation}
 As the eigenvalues are related to some energy scale (squared), this procedure acts as a low-pass filter in the sense of a Fourier decomposition: with lower values $N$ of included modes the Laplace filtering gets stronger. At this point it should be stressed that Laplace filtering is completely different from smearing, because it is based on extended objects, namely the eigenmodes, and does not locally modify the gauge links in contrast to smearing.

In this sense this method is quite similar to Dirac filtering \cite{Diracfilter}. However, Laplacian modes are chirality blind and do not obey an index theorem, which connects the zero-modes of the (chiral) Dirac operator to the topological charge.
From a computational point of view Laplace filtering is much cheaper than Dirac filtering (with good chirality properties).

After filtering we measure the topological charge density $q(x)=(1/16\pi^2) {\rm Tr}\{F_{\mu\nu}(x)\widetilde{F}^{\mu\nu}(x)\}$,
where $F_{\mu\nu}$ is an improved discretization of the field strength tensor \cite{Bilson-Thompson2003}, which combines $1\times1$, $2\times2$ and $3\times3$ loops to achieve $O(a^4)$-improvement at tree-level\footnote{for details on the improvement coefficients see \cite{Bilson-Thompson2003}}.

In order to get results for dynamical and quenched configurations which can be easily compared, we have chosen two $SU(3)$ ensembles with the same lattice spacing and the same physical volume (see Tab.~\ref{tab: configurations}). The ensembles were generated with the L\"uscher-Weisz gauge action and a chirally improved Dirac operator. For the dynamical simulations two flavors of mass degenerate light quarks were used (details can be found in \cite{CIRef}). 

\begin{table}[!t]
\centering
 \begin{tabular}{l @{\extracolsep{5mm}} c @{\extracolsep{5mm}}c @{\extracolsep{5mm}}c @{\extracolsep{5mm}}c }
\hline\hline
 & lat. size & lat. spacing & $\beta_{L W}$ \\ \hline
quenched & $16^3\cdot32$ &  0.148 & 7.90 \\ 
dynamical & $16^3\cdot32$ & 0.150 & 4.65 \\ \hline\hline 
\end{tabular}
\caption{Details on the gauge configurations used in this work.}\label{tab: configurations}
\end{table}

\section{Comparison of the Methods}\label{sec: Comparing the methods}
For quenched $SU(2)$ lattice configurations we had observed a similarity of the topological density obtained with the different filtering methods \cite{Bruckmann2007c}. For a quantitative comparison one needs a measure of the similarity. In \cite{Bruckmann2007c} the following quantity was introduced:
\begin{equation}\label{eq: Xi}
\Xi_{A B}\equiv \frac{ \chi_{A B}^2}{ \chi_{A A}\;\chi_{B B}}
\end{equation}
with $\chi_{A B}\equiv(1/V)\sum_x(q_A(x)-\overline{q}_A)(q_B(x)-\overline{q}_B)$ being the correlator of two topological charge densities $q_A(x)$ and $q_B(x)$ (mean values $\overline{q}=Q/V$ are subtracted for convenience).

$\Xi$ is a positive quantity and equals $1$, if the densities differ only by a constant scaling factor. The great advantage of this quantity is that we do not have to know the normalization factors of the topological charge densities, since these factors drop out in eq.~(\ref{eq: Xi}) and we are able to compare only the relative difference at each lattice point.

Moreover, we can define `best matching' pairs of filter parameters, by maximizing $\Xi_{A B}$. To be precise, we keep the smearing parameters constant and compute $\Xi_{A B}$ for up to 50 APEsteps and 50 Stout steps respectively up to 500 Laplace modes. More smearing steps have not been applied, as too much smearing will destroy the local topological structure, whereas more Laplace modes would be too expensive.

\begin{figure}
\centering
\subfigure[APE vs. Stout ]{\label{fig: Xi ape stout dynamical}
 \includegraphics[trim=10 0 10 0,width=0.34\textwidth]{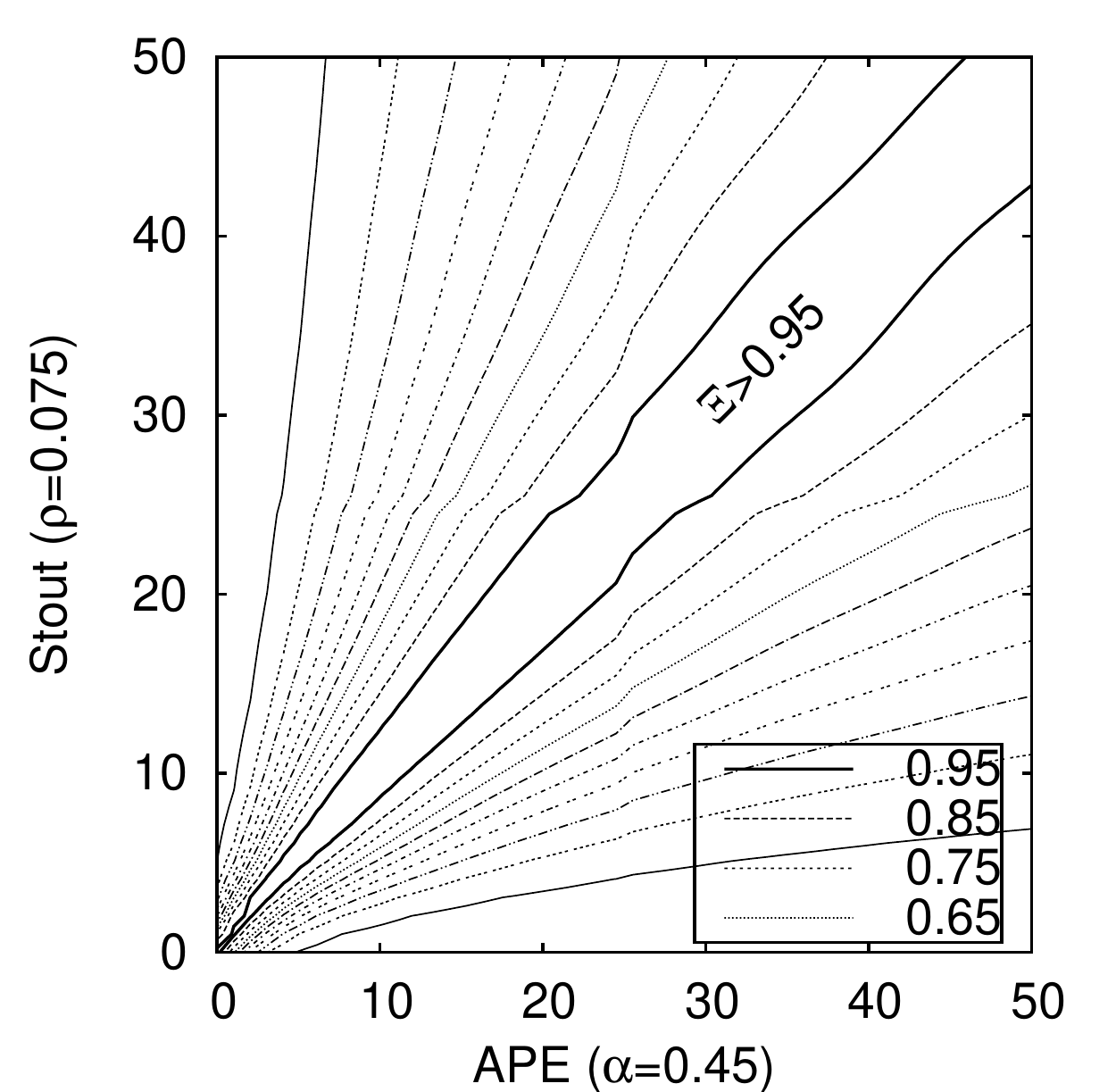}}
 \subfigure[APE vs. Laplace]{\label{fig: Xi ape laplace dynamical}
 \includegraphics[trim=10 0 10 0,width=0.34\textwidth]{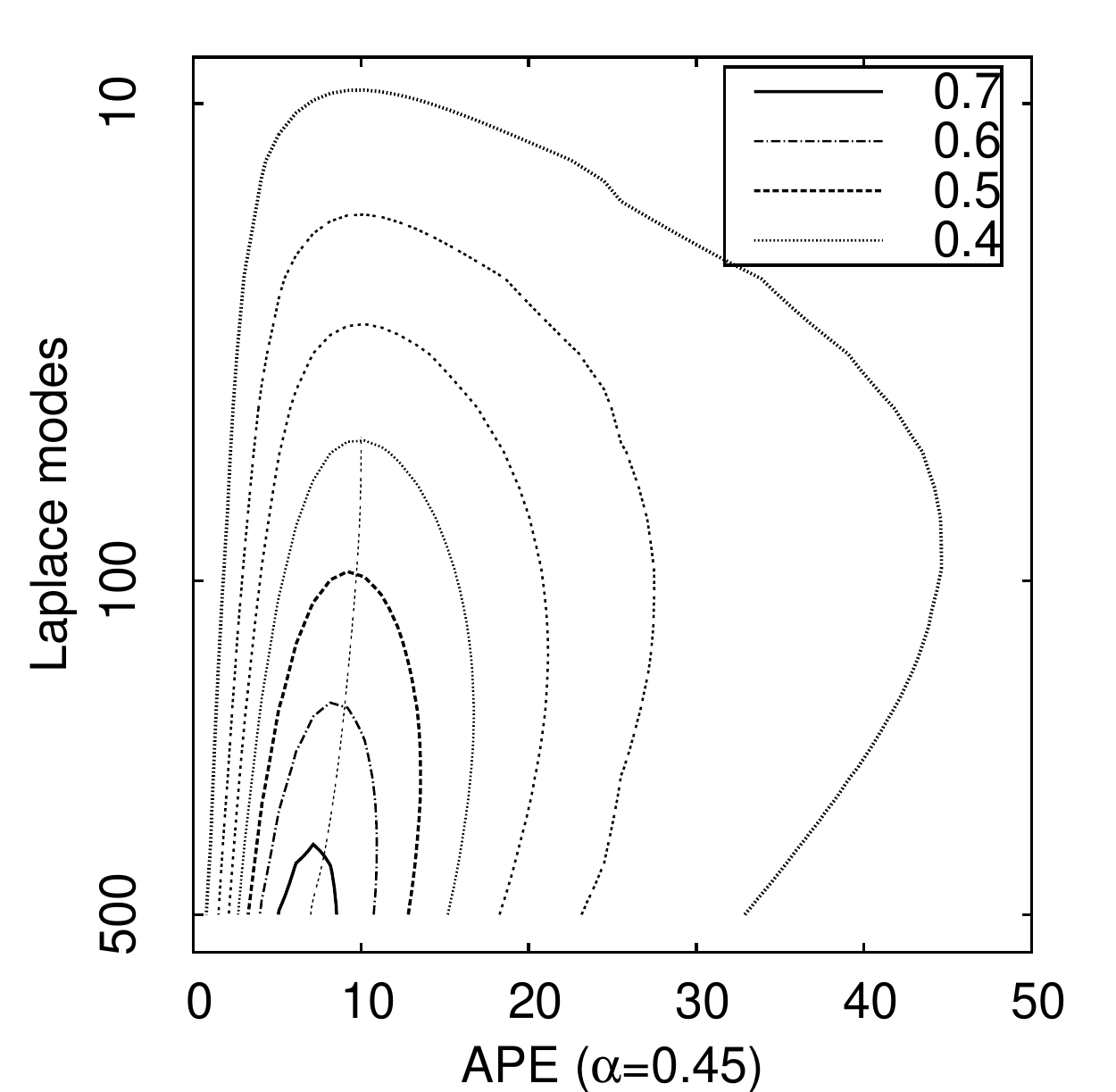}}
\caption{Comparison of APE with Stout smearing and APE smearing with Laplace filtering for an ensemble dynamical configurations. The degree of similarity $\Xi$ (see the definition (\protect\ref{eq: Xi})) is given by the contour lines. Shown are mean values of 80 configurations for APE vs. Stout smearing and 30 configurations for APE smearing vs. Laplace filtering. We find for APE vs. Stout smearing a band with $\Xi > 0.95$. Thus, both methods find the same structures and artifacts. The values of $\Xi$ for APE smearing vs. Laplace filtering does not reach such high values. For mild filtering we have plotted a ridge line of best matching parameters that will be used in the cluster analysis later. The described behavior applies also to quenched configurations .\label{fig: Xi}}
\end{figure}

The matching of APE and Stout smearing, see Fig.\ \ref{fig: Xi ape stout dynamical}, shows a one-to-one correspondence of the topological charge densities \cite{Bruckmann2009,Bruckmann2009c}. Thus, the different smearing techniques reveal almost the same structures, but they will also find the same unphysical artifacts. To control the ambiguities it is necessary to complement smearing by an independent filtering method, in our case Laplace filtering.

The comparison is not so perfect when we consider APE smearing and Laplace filtering, see Fig.~\ref{fig: Xi ape laplace dynamical}. Also here we find some kind of ridge line of the best matching values of $\Xi$. This ridge is quite pronounced for weak filtering, but we are confronted with the problem to identify the best matching parameters in the strong filtering regime, i.e. for less than 50 Laplace modes and more than 20 smearing steps. Furthermore, there is no set of best matching parameters with $\Xi>0.5$ in this regime and thus one can hardly say that both methods agree at all ($\Xi=0.5$ roughly means that only every second object is found by both filters). This finding flags a warning that any kind of smearing should only be applied with great care.

\section{Cluster Analysis of the Topological Charge Density}\label{sec: cluster analysis}

In order to obtain information about the topological structure of the QCD vacuum that can be compared to continuum models, we analyze the cluster structure of the topological charge density. Two lattice points belong to the same cluster, if they are nearest neighbors and have the same sign of the topological charge density.

 For such clusters we had found an interesting power law \cite{Bruckmann2007c,Bruckmann2009c}. To that end one cuts the absolute value of the topological charge at a (variable) cut-off $q_{\rm cut}$ and considers the number of clusters above this cut-off as a function of the fraction $f$ of all points obeying $|q(x)|>q_{\rm cut}$ w.r.t.\ the total number of lattice points.

The exponent $\xi\equiv {\rm d}\log N_{\rm clust}(q_{\rm cut})/{\rm d}\log N_{\rm points}(q_{\rm cut})$ of this power law is highly characteristic for the topological structure of the QCD vacuum. Different models lead to different predictions. This allows for a very sensitive test.

Pure noise, for instance, yields an exponent $\xi=1$, because every point forms its own cluster. On the other hand, the exponent is close to zero for very smooth densities with large structures, because in such a situation a small change in the cut-off will add new points but not reveal new clusters.

The exponent can be related to the size distribution $d(\rho)\sim \rho^\beta$ of equally charged topological objects with arbitrary shape function and arbitrary dimensionality \cite{Bruckmann2007c}. For instance a dilute gas of instantons would have $\xi=0.64$ in the quenched case and $\xi=0.66$ in the dynamical case ($N_f=2$). 

A cluster analysis has one great advantage. It allows to reduce  ambiguities coming from a single filter by taking only those clusters into account, which are common to different filters. If there is an artifact coming from one method, it is unlikely that this artifact will also be seen by the other one, such that the common structures are almost free of ambiguities.

\begin{figure}
\centering
\subfigure[Smearing]{\label{fig: APE cluster analysis}
  \includegraphics[width=0.437\textwidth]{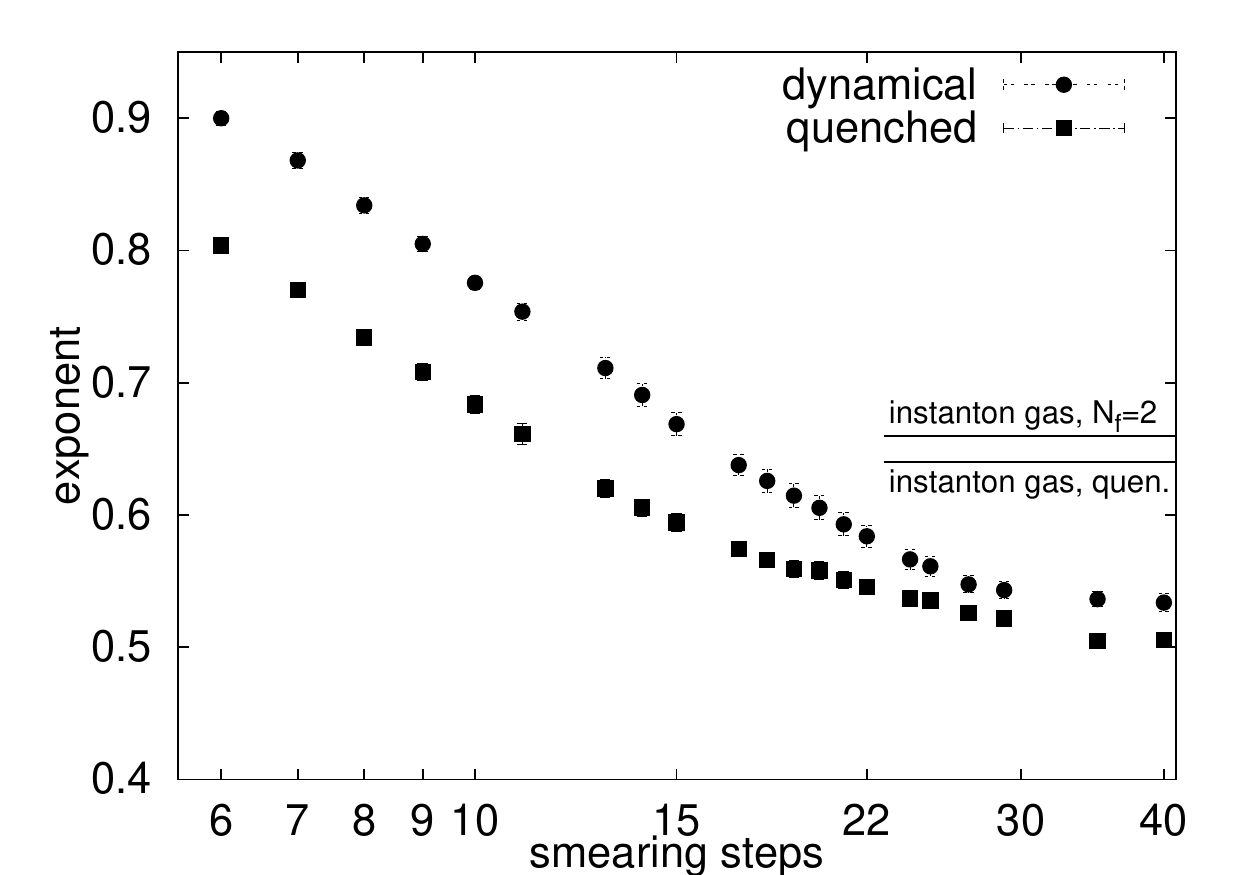}}
\subfigure[Laplace filtering]{\label{fig: Laplace cluster analysis}
 \includegraphics[width=0.437\textwidth]{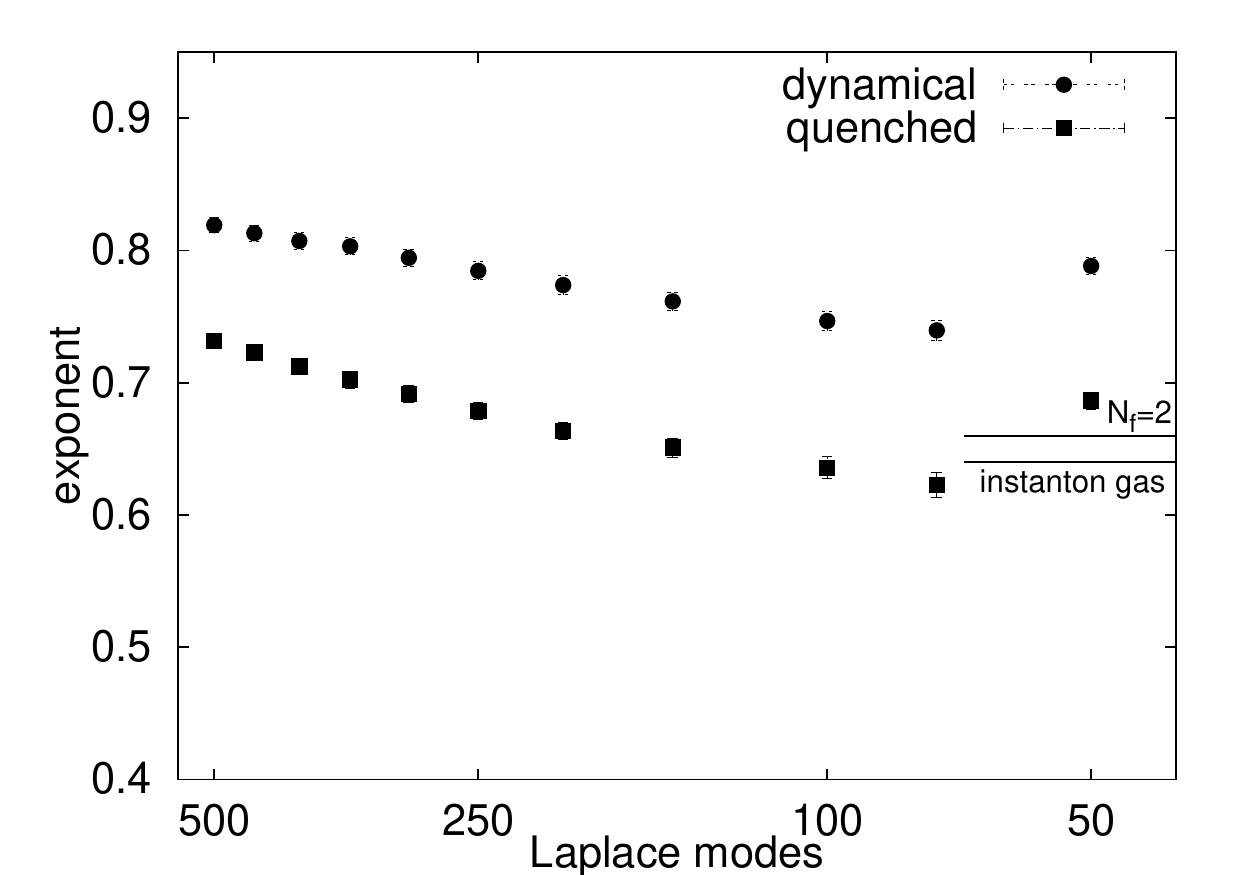}}
\subfigure[matched Filtering]{\label{fig:Laplace Ape matched cluster analysis}
\includegraphics[width=0.445\textwidth]{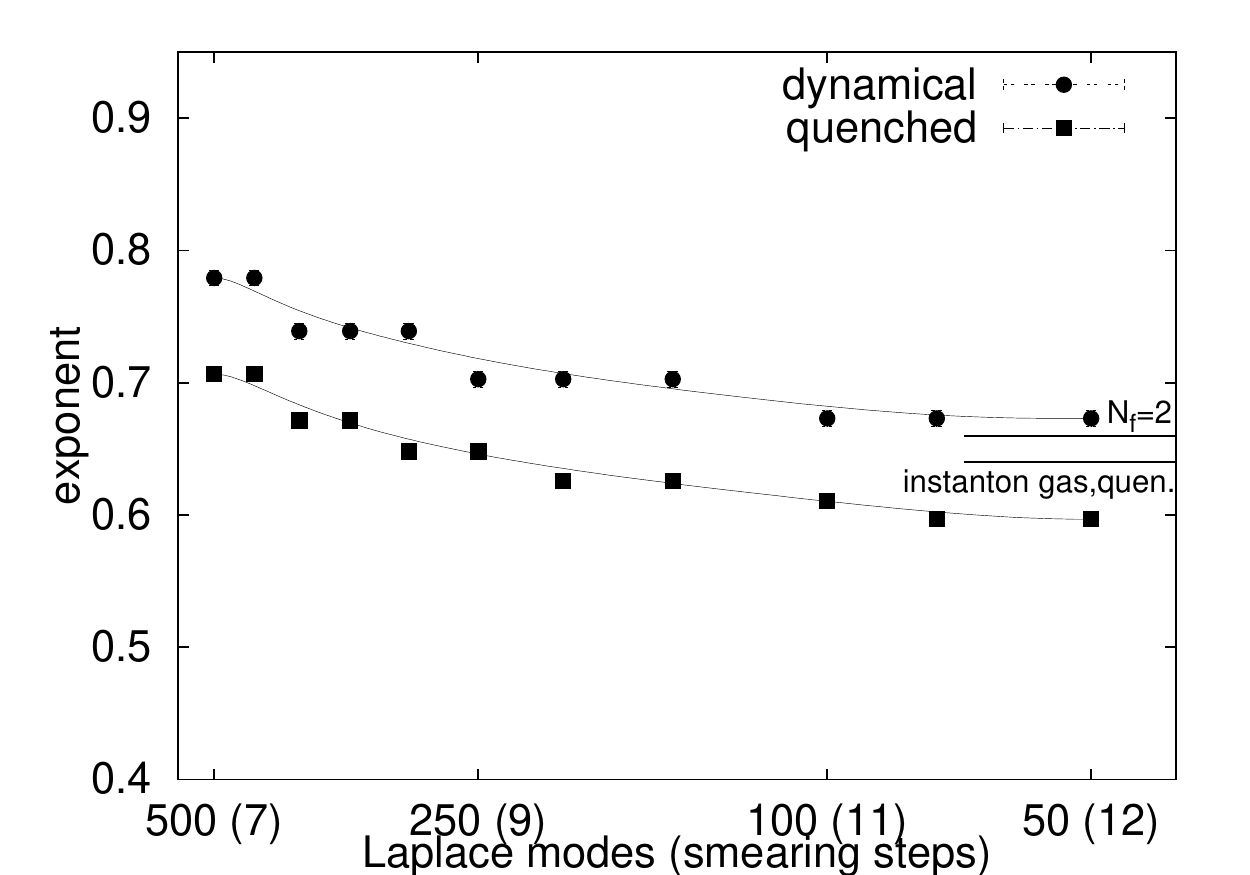}}
\caption{Exponent $\xi$ for the clusters found by APE smearing (a), Laplace filtering (b) and for a matched filtering of both methods (c). The exponents predicted by the dilute instanton gas have been included in all plots. In the latter plot the measured exponents have a step-like behavior, because the number of smearing steps is an integer quantity and we have to match several numbers of Laplace modes to the same number of smearing steps. The smooth lines are interpolations corresponding to a ``continuous'' smearing parameter. The number of smearing steps (in parentheses) refers to the matching of quenched configurations and is slightly different for the dynamical case. Errors from the ensemble average have been included, but are partly too small to be seen.}
\end{figure}
The exponent for clusters common to APE and Stout smearing can be found in Fig.\ \ref{fig: APE cluster analysis}. Obviously, the exponents of the dynamical configurations lie above the quenched values. The difference of the cluster exponents quenched vs.\ dynamical vanishes for stronger smearing ($\sim$ 30 steps) 
and the exponents settle down to almost the same plateau. So we have reasons to believe that too much smearing destroys the impact of dynamical quarks.

In Fig.\ \ref{fig: Laplace cluster analysis} the same analysis is done for Laplace filtering (only). We find a slightly bigger difference of the exponents of dynamical and quenched configurations. Furthermore, this difference remains, unlike for APE smearing, at every stage of filtering. At first sight the value for 50 modes seems to be odd, but this is just an artifact of the incomplete reconstruction of the topological background. For a small number of modes, we get a very spiky structure and we will find many new clusters, if we lower the cut-off. Thus, we get a higher exponent.

To get rid of the ambiguities of APE smearing and Laplace filtering, we perform in Fig.\ \ref{fig:Laplace Ape matched cluster analysis} a matched cluster analysis. i.e. we consider only those clusters, which are common to both methods. We use the optimal set of filter parameters according to the maximal values of $\Xi$ (and corresponding to points on the ridge line of Fig. \ref{fig: Xi}) on the abscissa.

Fig.\ \ref{fig:Laplace Ape matched cluster analysis} shows that the \emph{cluster exponent is larger in the dynamical case} than in the quenched case. According to the considerations from above, this 
implies a larger exponent $\beta$ of the size distribution. Hence, very small topological objects become suppressed when quarks are taken into account. 

In the plots we have included the values of the exponent in instanton gases. These have the same ordering with a slightly smaller difference. Generally our measured values of the exponent are not far off the instanton gas values (which was not the case in $SU(2)$ \cite{Bruckmann2007c}).

\begin{figure}
\centering
\includegraphics[trim= 0 0 30 0,width=0.43\textwidth]{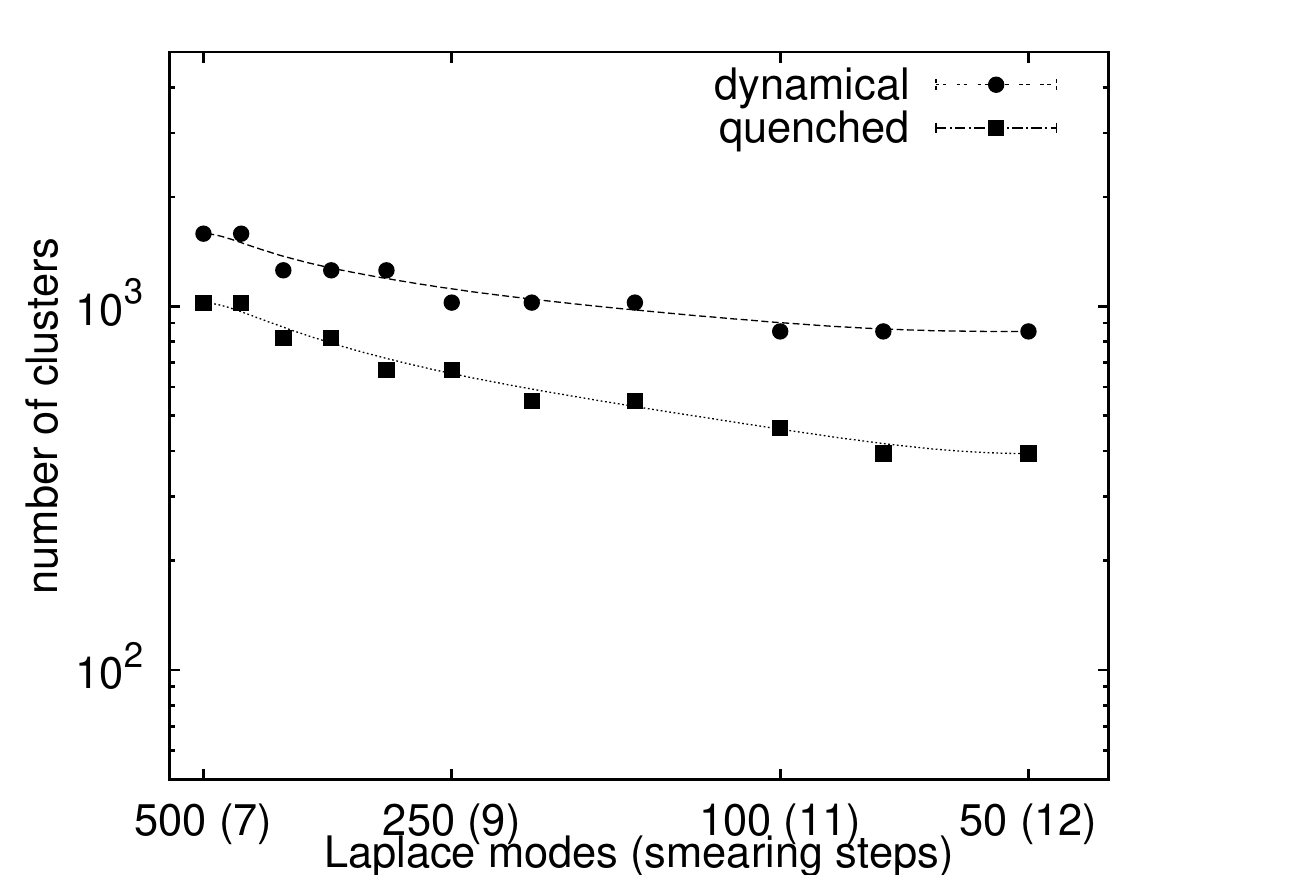}
\caption{Total number of distinct clusters for a constant fraction $f=0.0755$ of points lying above the cut-off with smooth interpolation.}\label{fig:absolute number of matched clusters}
\end{figure}
Another interesting quantity is the absolute number of clusters above a certain cut-off, see Fig.\ \ref{fig:absolute number of matched clusters}. In the dynamical case we find almost twice as many clusters. Hence, we observe a \emph{more fragmented topological structure in the presence of dynamical quarks}.

Our findings can be compared to the results obtained by the Adelaide group \cite{Moran2008a}. They have observed an ``increasing density of nontrivial field configurations'' and a suppression of small instantons.
Using a completely different approach, we come to the same conclusion. The advantages of our analysis are that our dynamical fermions have nicer chiral properties than the staggered ones. Moreover, we do not have to postulate any shape function of the topological objects. Therefore, our method works for all topological building blocks and with our matched analysis we can obtain results which are almost free from ambiguities of the filtering process.

\section{Conclusions}

We have performed a comparative study on the local topological structure of quenched and dynamical configurations seen by various filtering methods, namely APE smearing, Stout smearing and Laplace filtering.

To this end we have analyzed the power-law behavior of topological charge clusters seen by the individual methods and common to both methods.This analysis shows clearly a larger exponent in the presence of dynamical quarks, which can be related to a larger exponent of the size distribution $d(\rho)\sim \rho^\beta$ of topological objects. Hence, very small topological objects become suppressed in the dynamical case. Moreover, the total number of clusters for a constant total cluster volume is substantially higher for dynamical configurations. Thus the topological structure of the vacuum is more fragmented in the presence of fermions.

\begin{acknowledgments}
We would like to thank G. Bali, N. Cundy, E.-M. Ilgenfritz, M. M\"uller-Preussker and S. Solbrig for helpful comments and discussions. FB has been supported by DFG BR 2872/4-1 and FG by SFB TR-55.
\end{acknowledgments}

\end{document}